\documentclass{ifacconf}

\usepackage{graphicx}  
\usepackage{natbib}    
\usepackage{amsfonts,amsmath} 
\usepackage{color}
\usepackage{multirow}
\usepackage{flushend}
\usepackage{url}
\usepackage{verbatim}

\def\R{\ensuremath{\mathbb{R}}}

\def\O{\ensuremath{\mathcal{O}}}
\def\L{\ensuremath{\mathcal{L}}}

\begin{document}
\begin{frontmatter}
\title{Analysing control-theoretic properties of nonlinear synthetic biology circuits\thanksref{footnoteinfo}}

\thanks[footnoteinfo]{This work has been submitted to IFAC for possible publication.\\
Supported by grant PID2020-113992RA-I00 funded by MCIN/AEI/ 10.13039/501100011033, grant PID2023-146275NB-C21 funded by MICIU/AEI/10.13039/501100011033 and ERDF/EU, and grant ED431F 2021/003 funded by the Xunta de Galicia, Conseller\'ia de Cultura, Educaci\'on e Universidade.}

\author[UVIGO]{Ant\'on Pardo} 
\author[UVIGO]{Sandra D\'iaz Seoane} 
\author[UVIGO]{Dorin A. Ionescu} 
\author[OXFORD]{Antonis Papachristodoulou} 
\author[CITMAga,UVIGO]{Alejandro F. Villaverde}

\address[CITMAga]{CITMAga, 15782 Santiago de Compostela, Galicia, Spain}
\address[UVIGO]{Universidade de Vigo, Department of Systems and Control Engineering, 36310 Vigo, Galicia, Spain}
 \address[OXFORD]{Department of Engineering Science, University of Oxford, Oxford OX1 3PJ, UK}

\begin{abstract}  
Synthetic biology is a recent area of biological engineering, whose aim is to provide cells with novel functionalities. A number of important results regarding the development of control circuits in synthetic biology have been achieved during the last decade. A differential geometry approach can be used for the analysis of said systems, which are often nonlinear. Here we demonstrate the application of such tools to analyse the structural identifiability, observability, accessibility, and controllability of several biomolecular systems. We focus on a set of synthetic circuits of current interest, which can perform several tasks, both in open loop and closed loop settings. We analyse their properties with our own methods and tools; further, we describe a new open-source implementation of the techniques.  
\end{abstract}


\end{frontmatter}
%===============================================================================

%%%%%%%%%%%%%%%%%%%%%%%%%%%%%%%%%%%%%%%%%%%%%%%
\section{Introduction}

Feedback loops are ubiquitous in biological systems, both natural and artificial. Synthetic biology is an interdisciplinary research field concerned with ``re-programming'' cells, providing them with new or modified functionalities \citep{qian2018programming}. In this context, key goals are to achieve some desired dynamics and to reduce the effect of uncertainty. Since these systems are usually nonlinear, tools from nonlinear control theory are required for this aim.
%Synthetic biological circuits can be used to enforce some desired dynamical behavior. 
The theoretical ability to drive a system to a final state is given by its \textit{accessibility} and \textit{controllability} \citep{lewis2001brief}. Biological systems often have components with a high degree of uncertainty. To obtain a good characterization, it is important to be able to determine correctly the values of any unknown parameters and unmeasured variables. The possibility of performing these tasks successfully is given by a model's \textit{identifiability} and \textit{observability} \citep{chatzis2015observability,villaverde2019observability}. 

In this work we are interested in analysing such control theoretic properties from a \textit{structural} point of view, i.e., focusing on the constraints imposed by the equations that define system dynamics, rather than on \textit{practical} limitations introduced by measurement uncertainty \citep{wieland2021structural}. To model the systems under study we use nonlinear ordinary differential equations (ODEs), which are adequate for biomolecular systems as long as the number of molecules is sufficiently large. They are also helpful as approximate models of systems that are known to exhibit stochastic behaviour.
Under these assumptions, the aforementioned structural properties -- accessibility, controllability, structural identifiability and observability -- can be studied using a differential geometry approach \citep{hermann1977nonlinear}. 

As the field of synthetic biology moves from an initial ``modules area'' to a  ``systems era'' of increased complexity, the analysis of circuit properties becomes more imperative \citep{briat2020biology}.
According to \cite{qian2018programming}, many of the remaining challenges in synthetic biology can be addressed by a control-theoretic approach, but theory is currently lagging behind the development of biological tools. Likewise, \cite{baetica2019control} claim that control theory has yet to be fully applied to the understanding and engineering of biological systems. 
Indeed, there are few works reporting these types of analyses of synthetic biology systems, despite a few recent examples \citep{diaz2023controllability,haus2023structural}. 

In this paper we contribute to fill this gap. To this end, we apply techniques from the control theory of nonlinear systems to a set of recently presented models of synthetic biology circuits. We show how the analyses can be performed with different tools, and present a new Python implementation of our own accessibility and controllability code, which was previously available only in MATLAB. These implementations provide a counterpart to our MATLAB and Python toolboxes for structural identifiability and observability analysis (i.e. STRIKE-GOLDD, STRIKEpy). 

%The remainder of this paper is organised as follows: Section \ref{sec:methods} provides an overview of the theoretical concepts and techniques used in this work; Section \ref{sec:synth} describes the synthetic biological circuits that we analyse; Section \ref{sec:results} presents the results of the analyses; and Section \ref{sec:discussion} summarizes the conclusions.

%%%%%%%%%%%%%%%%%%%%%%%%%%%%%%%%%%%%%%%%%%%%%%%
\section{Theory and methods}\label{sec:methods}

\subsection{Notation and systems}

We consider nonlinear systems of %affine-in-inputs 
ordinary differential equations of the form

\begin{equation}\label{mod}
   M:\left\{\begin{aligned}
    \dot{x}(t) & = & f\left(u(t),x(t),\theta\right), \\
    y(t) & = & h\left(u(t),x(t),\theta\right),
\end{aligned}\right. 
\end{equation}
where $f$ and $h$ are analytic functions; $x(t)\in\R^{n_x}$ is the vector of state variables at time $t$; $u(t)\in\R^{n_u}$ is the input vector, which is assumed to consist of infinitely differentiable functions; $y(t)\in\R^{n_y}$, the output vector; $\theta\in\R^{n_\theta}$, the parameter vector. 
In the following, we may omit the dependence on time for ease of notation, i.e., we may simply write $x, y, u$.
A particular case of \eqref{mod} are input-affine systems, the dynamics of which can be written as follows: 
\begin{equation}\label{affine}
\dot x  =  f(x,\theta) \ + \ \Sigma _{i=1}^{n_u} \ u_i \ g_i(x,\theta)
\end{equation}

%--------------------------------
\subsection{Differential geometric concepts}

Differential geometric concepts for the analysis of nonlinear systems are described e.g. by \cite{sontag1982mathematical}.
Below we provide the definitions required in this work.

\subsubsection{Lie derivative:}
Given the model \eqref{mod}, the Lie derivative of the output function $h$ along the vector field $f$ is:
\begin{align*}
    &L_{f}h=\frac{\partial h}{\partial x} f+\frac{\partial h}{\partial u} \dot{u}.
\end{align*}
This is sometimes called the ``extended'' Lie derivative because it takes into account the effect of time-varying inputs.	
	
Setting $L^{0}_{f}h=h,$ the $i$-order extended Lie derivative of $h$ can be recursively computed as:
\begin{align*}\label{ext-lie}
    &L^{i}_{f}h=\frac{\partial L^{i-1}_{f}h}{\partial x} f+\sum_{j=0}^{i-1}\frac{\partial L^{i-1}_{f}h}{\partial u^{(j)}} u^{\left(j+1\right)}.
\end{align*}
where $u^{(j+1)}$ $(j\geq 0)$ stands for the $(j+1)$-th time derivative of $u.$

\subsubsection{Lie bracket:}
The Lie bracket of two vector fields $f, g$ is another vector field given by:
\begin{align*}
    &[f, g] = \frac{\partial g}{\partial x} f - \frac{\partial f}{\partial x} g,
\end{align*}
%Lie brackets are used for the analysis of accessibility and controllability of affine systems \eqref{affine}, as will be explained in Section \ref{sec:acc}.

\subsubsection{Lie algebra:}
A set of vector fields $\mathcal L$ is a Lie algebra when it is a linear subspace of vector fields (that is, $\alpha f + \beta g \in \mathcal L$ when $f,g\in \mathcal L $) and when the Lie bracket is well defined (i.e. $[f, g]  \in \mathcal L$ when $f,g\in \mathcal L $).
The Lie algebra generated by a family of vector fields $\mathcal{P}$ is the smallest Lie algebra containing $\mathcal{P}$, and it is written as $\mbox{LA} \left[ \mathcal P \right]$.

\subsubsection{Distribution:}
A distribution is a map $\mathcal S $ between each $x \in X$ and a subspace $\mathcal S (x) \subset \mathbb R ^n$. 

%--------------------------------
\subsection{Observability and structural local identifiability}

\subsubsection{Observability:} Conceptually, a state $x_i(\tau)$ (that is, the $i^{\text{th}}$ element of the state variables vector) is observable if it can be determined from the output $y(t)$ and the input $u(t)$ in an interval $t_0 \leq \tau  \leq t \leq t_f$, for a finite $t_f$. Otherwise, it is unobservable. A model is observable if all its states are observable. Observability is usually considered as a local property, i.e. an observable state can be distinguished from any other states in a neighbourhood, but possibly not from all distant states. Here we adopt this viewpoint and consider observability in the local sense, as we do with identifiability. This enables the use of differential geometric techniques.

\subsubsection{Structural local identifiability:} A parameter $\theta_i$ is structurally locally identifiable (SLI) if, for almost any parameter vector $\theta^{*} \in \mathbb{R}^{n_\theta}$, there is a neighbourhood $\mathcal{N}(\theta^{*})$ where the following condition holds \citep{distefano2015dynamic}:
\begin{equation}\label{eq:sli}
    \hat{\theta} \in \mathcal{N}(\theta^{*}) \hspace{0.3cm} \text{and} \hspace{0.3cm} y(t,\hat{\theta})= y(t, \theta^{*}) \Rightarrow \hat{\theta}_i = \theta^{*}_i 
\end{equation}
If \eqref{eq:sli} is not true in any neighborhood of $\theta^{*}$, the parameter $\theta_i$ is \textit{structurally unidentifiable} (SU). If  \eqref{eq:sli} is true for all model parameters, the model is said to be SLI as well, and SU otherwise.
We will use the acronym SIO to refer to Structural Identifiability and Observability.

\subsubsection{Structural local identifiability and observability (SIO):}
As noted by \cite{tunali1987new}, structural local identifiability can be treated as a particular case of observability by considering the parameters as state variables that happen to be constant, i.e. their dynamics are given by $\dot\theta_i=f(u(t),x(t),\theta)=0$. Thus, the SIO of a model can be evaluated using a version of the observability rank condition introduced by \cite{hermann1977nonlinear}.
Here we follow this approach, which we have previously implemented in the MATLAB toolbox STRIKE-GOLDD \citep{villaverde2016structural,diaz2022strike}.
The core idea is to build an observability-identifiability matrix, $O_I$, and calculate its rank.
To this end, we augment the state vector as $\tilde x = [x,\theta],$ defining $n_{\tilde x} = n_x + n_\theta.$ The observability-identifiability matrix $O_I$ of a model \eqref{mod} is:
\begin{equation}\label{obsmat2}
   \O_I(\tilde x)=\left(\begin{aligned}
    \frac{\partial}{\partial \tilde x}h(\tilde x,u)\quad \;\;\\
    \frac{\partial}{\partial \tilde x}\left(\L_f h(\tilde x,u)\right)\;\;\\
    \vdots\qquad\quad\\
    \frac{\partial}{\partial \tilde x} \left(\L_f^{n_{\tilde x}-1}h(\tilde x,u)\right)
\end{aligned}\right).
\end{equation}
\eqref{mod} is SLI and observable if $\text{rank}(\O_I(\tilde x))=n_{\tilde x}.$ If $\text{rank}(\O_I(\tilde x))<n_{\tilde x},$ there is at least one unobservable variable and/or one unidentifiable parameter. Since the $i^{\text{th}}$ column of $\O_I(\tilde x)$ represents the partial derivative with respect to the $i^{\text{th}}$ element of $\tilde x,$ the SIO of an individual variable, $\tilde x_i,$ can be determined by removing the $i^{\text{th}}$ column and recalculating the rank. If the rank decreases, $\tilde x_i$ is observable (or SLI, if it is a parameter); if the rank remains unchanged, $\tilde x_i$ is observable (or SU).

%--------------------------------
\subsection{Accessibility and controllability}\label{sec:acc}

%Briefly, controllability is the property by which a dynamic system can be guided from an initial state to any point in its proximity. Accessibility is a less restrictive characteristic than controllability. 
%An \textit{ admissible control} $u:[0,T] \rightarrow U$ is any measurable function $u(t)$ giving rise to an absolutely continuous state trajectory $x(t)$ satisfying \eqref{mod} almost everywhere. 

\subsubsection{Reachable set:}
The set of all points $x_f=x(t)$ with $t\leq T$ that a system can reach from an initial point $x_0$ in time at most $T$ is called the reachable set:       
$$ \mbox{ Reach} \left( M, \ \leq T, \ x_0 \right) \ = \ 
         \bigcup_{0 \leq t \leq T}  \ \mbox{ Reach} 
            \left( \Sigma, \ t, \ x_0 \right)$$

\subsubsection{Accessibility:}
The system $M$ \eqref{mod} has the accessibility property from $x_0 \in X$ if for every $T>0$ the set $\mbox{ Reach} \left( M, \ \leq T, \ x_0 \right)$ has a nonempty (full dimensional) interior \citep{sussmann1987general}.

\subsubsection{Controllability:}
The system $M$ \eqref{mod} is small-time locally controllable (STLC) from $x_0 \in X$ if for every $T>0$ the set $\mbox{Reach}  \left( M, \ \leq T, \ x_0 \right)$ contains $x_0$ in its (non-empty) interior.

To analyse accessibility and controllability we adopt the methodology described by \cite{diaz2023controllability}, which was originally implemented in MATLAB. To widen its adoption we have developed a new version in Python. We provide both implementations as open source software 
({\small\url{https://github.com/afvillaverde/NLcontrollability}}). 

The test for accessibility is based on determining whether certain distributions defined by the Lie algebras generated by the vector fields of a system are full-dimensional. We use the Lie Algebraic Rank Condition (LARC) described by \cite{diaz2023controllability}, which provides a sufficient and necessary condition for accessibility. For controllability we consider the General Sufficient Condition (GSC). These tests are applicable to systems of the form \eqref{affine}, i.e. which are affine in the inputs. 
%We consider the `ARC' and `LARC' tests described by \cite{diaz2023controllability}; the former is a sufficient condition for accessibility, while the latter is sufficient and necessary. Likewise, for controllability we consider the sufficient conditions `LC' and `GSC'. These tests are applicable to systems of the form \eqref{affine}, i.e. which are affine in the inputs. 

%\textcolor{blue}{[REMOVE?] The aforementioned approach is not applicable to nonlinear systems \eqref{mod} that are not affine in the inputs. For these systems we use instead a test inspired by the analysis of input observability, as described in \textcolor{red}{(add reference to the paper when it is available).} It is based on the calculation of Lie derivatives instead of Lie brackets, and hence we refer to it as LDC (Lie Derivatives Condition).}

%%%%%%%%%%%%%%%%%%%%%%%%%%%%%%%%%%%%%%%%%%%%%%%
\section{Control circuits in synthetic biology}\label{sec:synth}

\begin{figure*}
\begin{center}
\includegraphics[width=17cm]{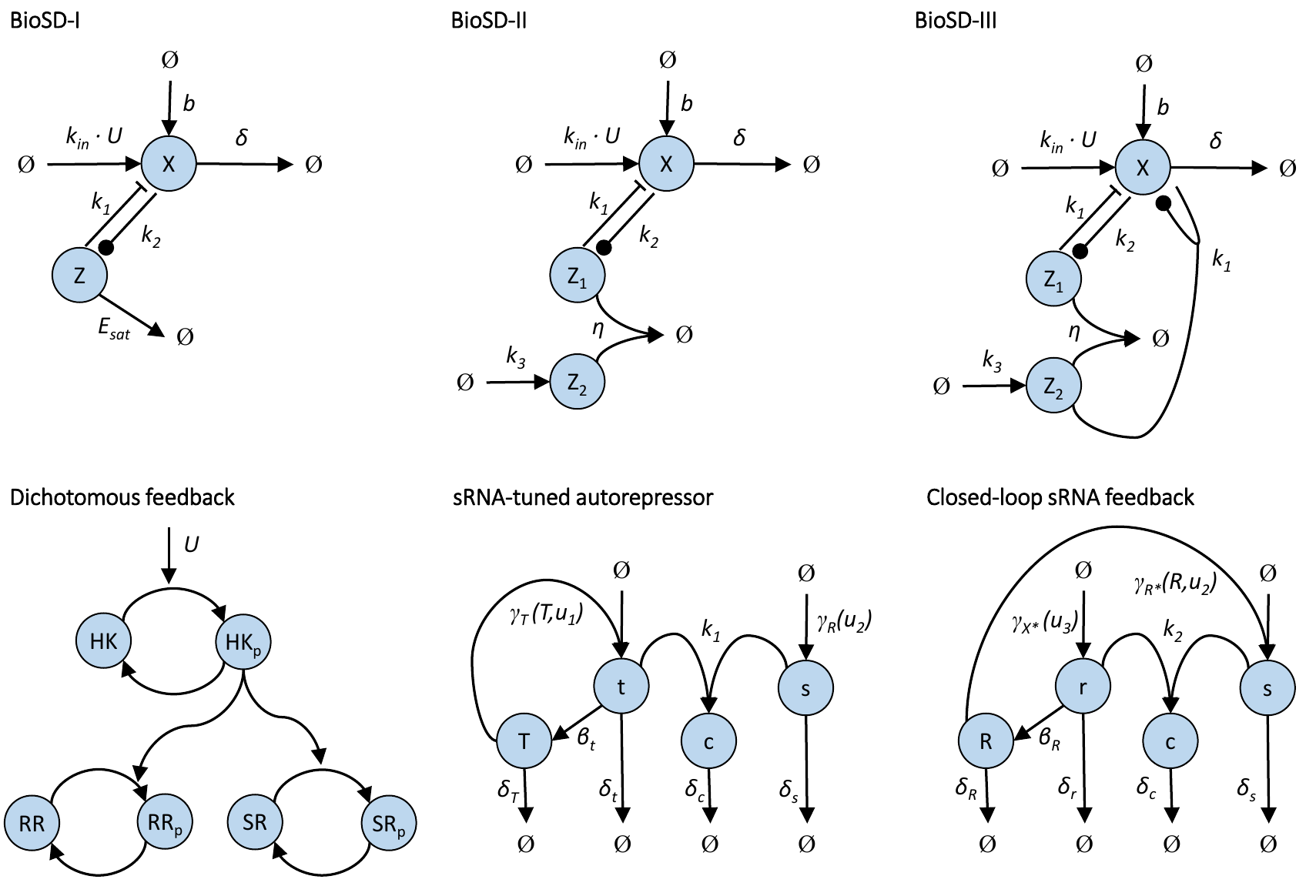}    % The printed column width is 8.4 cm.
\caption{Diagrams of the synthetic biology circuits analysed in this work.} 
\label{fig:circuits}
\end{center}
\end{figure*}

\cite{baetica2019control} and \cite{qian2018programming} reviewed control theoretic concepts of relevance to synthetic biology. %, such as modularity, homeostasis, and the trade-offs involved in stability, performance, and robustness. %Recent developments include antithetic feedback \citep{briat2016antithetic}...
Importantly, biological systems are typically positive, a property that has implications for their control \citep{briat2020biology}. However, this feature is not an obstacle for obtaining signals that can in principle be negative, such as derivatives, since this goal can be achieved by adding a bias \citep{alexis2021biomolecular}; likewise, it is not an obstacle for performing the analyses, as noted by \cite{diaz2023controllability}.
\cite{haus2023structural} analysed the structural identifiability of several models of biomolecular controller motifs, classified either as `basic' or `antithetic'. In this paper we focus on a set of circuits of which we analyse their accessibility, controllability (when possible), structural identifiability and observability. We provide their mathematical description in the remainder of this section, and their topologies in Fig. \ref{fig:circuits}. 

%--------------------------------
\subsection{Molecular topologies for signal differentiation}

\cite{alexis2021biomolecular} presented three topologies that perform signal differentiation. They can serve several purposes, such as acting as speed biosensors or implementing derivative control actions. We refer to them as BioSD-I, BioSD-II and BioSD-III, and give their equations below.
We use a generic notation for their description, where the parameters are written as $p_i$, and the derivative signal is $x_1(t)\approx \dot u(t).$ 

\subsubsection{BioSD-I:}
\begin{center}
    \[ \dot{X} = k_{in} \cdot U + b - k_{1} \cdot X \cdot Z - \delta \cdot X \]
    \[ \dot{Z} = k_{2} \cdot X - k_{3} \]
\end{center}

\subsubsection{BioSD-II:}
\begin{center}
    \[ \dot X = k_{in} \cdot U + b - k_{1} \cdot X \cdot Z_{1} - \delta \cdot X \]
    \[ \dot{Z}_1 = k_{2} \cdot X - \eta \cdot Z_{1} \cdot Z_{2} \]
    \[ \dot{Z}_2 = k_3 - \eta \cdot Z_{1} \cdot Z_{2} \]
\end{center}

\subsubsection{BioSD-III:}
\begin{center}
    \[ \dot X = k_{in} \cdot U + b - k_{1} \cdot X \cdot Z_{1} + k_{1} \cdot X \cdot Z_{2} - \delta \cdot X  \]
    \[ \dot{Z}_1 = k_{2} \cdot X - \eta \cdot Z_{1} \cdot Z_{2} \]
    \[ \dot{Z}_2 = k_{3} - \eta \cdot Z_{1} \cdot Z_{2} \]
\end{center}

Additionally, \cite{alexis2021biomolecular} introduced a more realistic version of BioSD-II, in which the activation of $x_2$ by $x_1$ takes place with Michaelis-Menten kinetics:

\subsubsection{BioSD-II-MM-simple:}
\begin{center}
    \[ \dot X = k_{in} \cdot U + b - k_{1} \cdot X \cdot Z_{1} - \delta \cdot X \]
    \[ \dot{Z}_1 = \frac{V_{max} \cdot X}{X+K_m} - \eta \cdot {Z}_1 \cdot {Z}_2 \]
    \[ \dot{Z}_2 = k_{3} - \eta \cdot Z_{1} \cdot Z_{2} \]
\end{center}

\subsubsection{BioSD-II-MM-complex:}
\begin{center}
    \[ \dot X = k_{in} \cdot U + b - k_{1} \cdot X \cdot Z_{1} - (\delta+\gamma)\cdot X \]
    \[ \dot{Z}_1 = \frac{V_{max} \cdot X}{X+K_m} - \eta \cdot Z_{1} \cdot Z_{2}-\gamma \cdot Z_{1} \]
    \[ \dot{Z}_2 = k_{3} - \eta \cdot Z_{1} \cdot Z_{2}- \gamma \cdot Z_{2}\]
\end{center}

%--------------------------------
\subsection{Dichotomous Feedback}\label{sec:dichotomous}

Natural biological systems may exhibit dichotomous feedback, which works through sequestration of a molecule or a signal. \cite{sootla2022dichotomous} proposed several ways of implementing this functionality. Here we study the following model, which is described in equations (2.7) of their article:
\begin{align*}
    \dot{HK} &= \beta_{HK} - \delta \cdot HK - k_{ap}(I) \cdot HK \\
    &+ k_{t} {\cdot} \left(\frac{\beta_{HK}}{\delta} - HK\right) {\cdot}  RR + k_{tc} {\cdot} \left(\frac{\beta_{HK}}{\delta} - HK\right) {\cdot} SR \\ %<- trick: writing {\cdot} instead of \cdot reduces horizontal spacing
    \dot{RR} &= \beta_{RR} - \delta \cdot RR - k_{t} \cdot \left(\frac{\beta_{HK}}{\delta} - HK\right) \cdot RR\\ &+ k_{p} \cdot HK \cdot \left(\frac{\beta_{RR}}{\delta} - RR\right)\\
    \dot{SR} &= \beta_{SR} - \delta \cdot SR - k_{tc} \cdot \left(\frac{\beta_{HK}}{\delta} - HK\right) \cdot SR\\ &+ k_{pc} \cdot HK \cdot \left(\frac{\beta_{SR}}{\delta} - SR\right)\\
\end{align*}
where $k_{ap}(I)=k_{ap-max}\frac{I}{I+K_{da}},$ and $I$ is the input signal. Note that taking $k_{ap}$ as the input signal yields a model that is affine in the inputs. We will also analyse this model considering the production rates $\beta_{*}$ as inputs that can be modified.

\begin{comment}
We also analyse the next model, described in equations (2.8) of \citep{sootla2022dichotomous}:
\begin{align*}
    \dot{HK} = &\beta_{HK} - \delta \cdot HK - k_{ap}(I) \cdot HK\\ & + k_{t} {\cdot} \left(\frac{\beta_{HK}}{\delta} - HK\right) {\cdot} RR+ k_{tc} {\cdot} \left(\frac{\beta_{HK}}{\delta} - HK\right) {\cdot} SR \\
    \dot{RR} = &\beta_{RR} - \delta \cdot RR - k_{t} \cdot \left(\frac{\beta_{HK}}{\delta} - HK\right) \cdot RR\\ &+ k_{p} \cdot HK \cdot \left(\frac{\beta_{RR}}{\delta} - RR\right)\\
    \dot{SR} = &\beta_{SR} - \delta \cdot SR - k_{tc} \cdot \left(\frac{\beta_{HK}}{\delta} - HK\right) \cdot SR\\ &+ k_{pc} \cdot HK \cdot \left(\frac{\beta_{SR}}{\delta} - SR\right)\\
\end{align*}
\end{comment}

%--------------------------------
\subsection{Negative Feedback}\label{sec:negative}

We consider two synthetic circuits based on engineered small RNAs (sRNAs) presented by \cite{kelly2018synthetic}. % These models have expressions that are not affine in the inputs. 

\subsubsection{sRNA-tuned autorepressor:} % Detailed model (equations 1, 3 OR 4, 8 of the Supplementary Information of \citep{kelly2018synthetic}; ``Kelly-1''):
\begin{center}
    \[ \dot{t} = \gamma_{T}- \delta_{t} \cdot t - K_{1} \cdot t \cdot s \]
    \[\dot{s} = \gamma_{R} - \delta_{s} \cdot s- K_{1} \cdot t \cdot s\]
    \[ \dot{c} = K_{1} \cdot t \cdot s- \delta_{c} \cdot c \]
    \[ \dot{T} = \beta_{T} \cdot t - \delta_{T} \cdot T \]
\end{center}
Where $\gamma_{T}$ and $\gamma_{R}$ are defined as follows:

\begin{align*}
\gamma_T&=\frac{\alpha_{t}}{1+\left(\frac{T}{K_{T}\left(1+\left(u_1 / K_{u_1}\right)^{n_{u_1}}\right)}\right)^{n_{T}}}\\ &+\frac{\alpha_{L}\left(\frac{T}{K_{T}\left(1+\left(u_1 / K_{u_1}\right)^{n_{u_1}}\right)}\right)^{n_{T}}}{1+\left(\frac{T}{K_{T}\left(1+\left(u_1 / K_{u_1}\right)^{n_{u_1}}\right)}\right)^{n_{T}}}
\end{align*}
\begin{center}
\begin{equation}
\gamma_R=\alpha_{r} \frac{u_2}{K_{u_2}+u_2}
\end{equation}
\end{center}
The states are: $t$, mRNA concentration; $s$, sRNA concentration; $c$, sRNA-mRNA complex concentration; and $T$, TetR-GFP complex concentration.
Similarly to the models in Section \ref{sec:dichotomous}, this model is only affine in the inputs if these are taken to be $\gamma_T$ and $\gamma_R$ instead of $u_1$ and $u_2.$ The same applies to the following model. 

\subsubsection{Closed-loop sRNA Feedback Circuit:} % Detailed model (equations 2, 4, 15, ``Kelly-2''):
\begin{center}
    \[ \dot{r} = \overset{*}{\gamma_{X}} - \delta_{r} \cdot r - K_{2} \cdot r \cdot s \]
    \[ \dot{s} = \overset{*}{\gamma_{R}}-\delta_{s} \cdot s - K_{2} \cdot r \cdot s \]
    \[ \dot{c} = K_{2} \cdot r \cdot s - \delta_{c} \cdot c \]
    \[ \dot{R} = \beta_{R} \cdot r - \delta_{R} \cdot R \]
\end{center}
Where $\overset{*}{\gamma_{X}}$ and $\overset{*}{\gamma_{R}}$ are defined as follows:
\begin{center}
\begin{equation}
\gamma_X^*=\alpha^*_{X} \frac{u_3}{K_{u_3}+\left(u_3\right)}
\end{equation}
\begin{equation}
\gamma_R^*=\alpha^*_{r} \frac{\left(\frac{R\cdot u_2}{K^*_{u_2}+u_2}\right)}{K_{R}+\frac{R \cdot u_2}{K^*_{u_2}+u_2}}
\end{equation}
\end{center}
The new states are $r$, mRNA concentration; and $R$, RhaS-GFP complex concentration.

%%%%%%%%%%%%%%%%%%%%%%%%%%%%%%%%%%%%%%%%%%%%%%%
\section{Results}\label{sec:results}

%--------------------------------
\subsection{Accessibility and controllability}

\setlength{\tabcolsep}{0.1cm}
\begin{table}[htb]
    \begin{center}
    \caption{Results: accessibility, controllability.}\label{tb:acc}
        \begin{tabular}{|c|c|c|c|}
            \hline
            \textbf{Case study} &  \textbf{Eq. point} & \textbf{Accessible} & \textbf{Controllable} \\
            \hline
            BioSD-I &  param. &  yes &  yes  \\
            \hline
            BioSD-II &  param. &  yes &  yes  \\
            \hline
            BioSD-II-MM-simple &  param. &  yes &  yes  \\
            \hline
            BioSD-II-MM-complex & not found &  yes$^{\text{ (neq)}}$ &  NA  \\
            \hline
            BioSD-III & param. &  yes &  yes  \\
            \hline
            Dichotomous Feedback & param.  & yes &  yes  \\
            \hline
            \begin{tabular}{c}sRNA-tuned auto-\\repressor (input: $\gamma_R$)\end{tabular} & param.  & yes &  yes\\
            \hline
            \begin{tabular}{c}Closed-loop sRNA \\ (inputs: $\gamma_R^*,\ \gamma_X^*$) \end{tabular}& param. & yes & yes  \\
            \hline
        \end{tabular}
    \end{center}
\end{table}
\setlength{\tabcolsep}{6pt}

%As a preliminary check of correctness, we applied our new Python implementation of the methodology to the set of case studies analysed in \citep{diaz2023controllability}. The results (not shown) were consistent with the ones reported in the aforementioned publication, this validating the new tool. 

%Next, we used the tool to analyse the systems described in Section \ref{sec:synth}.
Table \ref{tb:acc} shows the list of biosystems along with the results produced by the LARC and GSC tests, which assess accessibility and small-time local controllability, respectively. The GSC is not applicable (NA) in non-equilibrium points, and it cannot test if the model is inaccessible. In principle, tests are performed at the equilibrium (in the table, ‘param.’ means that the equilibrium point depends on the value of the parameters and inputs). When an equilibrium is not found, the tests are performed around a point $x_{i}$ = 1,  i= 1, … , $n_{x}$; this is denoted as $^{\text{(neq)}}$. 

We analysed the negative feedback models of Section \ref{sec:negative} by treating the $\gamma$ functions as inputs. Alternatively, one could consider as inputs the $u_*$ variables; however, doing so would make these models non-affine in the inputs, which would prevent the application of the accessibility and controllability tests.

%--------------------------------
\subsection{Structural local identifiability and observability}

Since identifiability and observability (SIO) depend on the model outputs, we have considered several possible output configurations for each system. First, we considered the output defined in the original publications. Additionally, we explored combinations of several states, always keeping the original output among the measured states. Thus we are able to assess how measurement availability influences the SIO of the unmeasured states and parameters. Table \ref{tb:sio} summarizes the results of these analyses. 

We analyse two versions of the Dichotomous Feedback circuit: one that replaces $k_{ap}(I)$ with $k_{ap-max}\frac{I}{I+K_{da}}$ in the equations, which we denote with ($I$), and another one which uses directly $k_{ap}.$ In each of these versions we perform the analyses when said variable is the input, and also when each of the production rates $\beta_*$ is the input (when a $\beta_*$ is not an input, it is treated as a parameter).
Likewise, we consider similar variants of the sRNA models, i.e., taking either the $\gamma$ or the $u$ variables as inputs.
%The SIO analysis of the models described in Section \ref{sec:negative} could not be performed with STRIKE-GOLDD due to the high number of Lie derivatives needed by the calculations, which consume too much memory. \textcolor{blue}{Instead, we have used STRUCID / GRAMIANS ...}
We note that the sRNA-tuned autorepressor model could only be analysed by assuming that at least two states can be measured; analyses with only one output required too much memory. 

The SIO results inform about the possibility of identifying the parameters for every possible output configuration. Let us consider for example the closed-loop sRNA circuit. A typical choice could be to take measurements on the RhaS-GFP complex, i.e. $R.$ As shown in Table \ref{tb:sio}, this would make it impossible to identify the sRNA-mRNA binding strength ($K_2$), the translation rate of RhaS-GFP mRNA ($\beta_R$), the degradation rate of the mRNA-sRNA complex ($\delta_c$), and the transcription rates $\alpha^{*}_r$ and $\alpha^{*}_X$. Furthermore, it would be impossible to infer any of the unmeasured state variables. In contrast, if one measures not only $R$ but also $c$, which is the sRNA-mRNA complex concentration, all the parameters become identifiable, and all state variables become observable.
Similar insights can be extracted for each of the output configurations shown in the aforementioned table. 

\begin{table*}[htb]
\begin{center}
    \caption{Results: structural identifiability, observability. For each model we consider different choices of inputs (u) and outputs (y); a slash (e.g. $A$/$B$) indicates that the result holds when either variable ($A$ or $B$) is taken as an input or output.
    Parameters are classified as `Identifiable' or `Non-identifiable', and states as observable (`Obs') or non-observable (`Non-obs'). 
    }\label{tb:sio}
    \begin{tabular}{|c|c|c|c|c|c|c|}
        \hline
        \textbf{Model}    & 
        \textbf{u} & \textbf{y}  & \textbf{Identifiable} & \textbf{Non-identifiable} & \textbf{Obs.}   & \textbf{Non-obs.} \\ \hline
         \multirow{3}{*}{BioSD I}         & 
        \multirow{3}{*}{U} & $X$     & $k_{in}$ b & $k_1$ $k_2$ $k_3$ $\delta$ & X    & Z  \\ \cline{3-7}
        & 
         & $Z$     & $k_1$ $k_3$ $\delta$ & $k_2$ $k_{in}$ b & Z    & X     \\ \cline{3-7}
                & 
         & all   & all               & -              & all   & -      \\ \hline
        \multirow{4}{*}{BioSD II}         & 
        \multirow{4}{*}{U} & $X$   & $k_{in}$ b $\delta$ & $k_1$ $k_2$ $k_3$ $\eta$    & X    & $Z_1$ $Z_2$  \\ \cline{3-7}
                 & 
         & $Z_1$/$Z_2$/($Z_1$ $Z_2$)   & $k_1$ $k_3$ $\delta$ $\eta$ & $k_2$ $k_{in}$ b & $Z_1$ $Z_2$ & X \\ \cline{3-7}
                 &   & X $Z_1$ & \multirow{2}{*}{ all}              & \multirow{2}{*}{ -  }            & \multirow{2}{*}{ all  } & \multirow{2}{*}{ - }     \\ \cline{3-3}
                 & 
         & $X$ $Z_2$ &                &               &    &       \\  \hline
        \multirow{4}{*}{\begin{tabular}{c}BioSD II MM\\ simple\end{tabular}}  & 
        \multirow{4}{*}{U} & X    & $k_{in}$ b $\delta$ $K_m$       & $k_1$ $k_3$ $\eta$ $V_{max}$    & X    & $Z_1$ $Z_2$  \\ \cline{3-7}
          & 
         & $Z_1$/$Z_2$/($Z_1$ $Z_2$)  & $k_1$ $k_3$ $\delta$ $\eta$ $V_{max}$ & $k_{in}$ b $K_m$ & $Z_1$ $Z_2$ & X     \\ \cline{3-7}
          &  & $X$ $Z_1$ & \multirow{2}{*}{all}               & \multirow{2}{*}{-}              & \multirow{2}{*}{all}   & \multirow{2}{*}{-}      \\ \cline{3-3}
          & 
         & $X$ $Z_2$ &   &   &  &   \\   \hline
        \multirow{4}{*}{\begin{tabular}{c}BioSD II MM\\ complex\end{tabular}} & 
        \multirow{4}{*}{U} & $X$    & $k_{in}$ b $\delta$ $K_m$  $\gamma$   & $k_1$ $k_3$ $\eta$ $V_{max}$   & X    & $Z_1$ $Z_2$  \\ \cline{3-7}
           & 
         & $Z_1$/$Z_2$/($Z_1$ $Z_2$)  & $k_1$ $k_3$ $\delta$ $\eta$ $V_{max}$ $\gamma$ & $k_{in}$ b $K_m$ & $Z_1$ $Z_2$ & X \\ \cline{3-7}
           & 
         & $X$ $Z_1$ & \multirow{2}{*}{all }              & \multirow{2}{*}{- }             & \multirow{2}{*}{all}   & \multirow{2}{*}{- }     \\ \cline{3-3}
           & 
         & $X$ $Z_2$ &              &             &    &       \\  \hline
        \multirow{4}{*}{BioSD III} & 
        \multirow{4}{*}{U} & X    & $k_{in}$ b & $k_1$ $k_2$ $k_3$ $\delta$ $\eta$ & X  & $Z_1$ $Z_2$  \\ \cline{3-7}
          & 
         & $Z_1$/$Z_2$/($Z_1$ $Z_2$) & $k_1$ $k_3$ $\delta$ $\eta$ & $k_2$ $k_{in}$ b & $Z_1$ $Z_2$ & X \\ \cline{3-7}
          & 
         & $X$ $Z_1$ & \multirow{2}{*}{all}               & \multirow{2}{*}{- }             & \multirow{2}{*}{all}   & \multirow{2}{*}{- }     \\ \cline{3-3}
          & 
         & $X$ $Z_2$ &               &             &    &    \\  \hline
         
        \multirow{3}{*}{\begin{tabular}{c}Dichotomous\\ Feedback ($I$)\end{tabular}} & $I$ & HK/RR/SR & all  & - & \multirow{2}{*}{all} & \multirow{2}{*}{-}  \\ \cline{2-5}
        & $\beta_{HK}$/$\beta_{RR}$/$\beta_{SR}$ & HK/RR/SR & \begin{tabular}{c}
           $\delta$ $k_t$ $k_{tc}$ $k_p$ $k_{pc}$\\
           $\beta_{*}$ 
        \end{tabular}        & $I$ $k_{ap-max}$ $K_{da}$ &     &   \\ \hline
        \begin{tabular}{c}Dichotomous\\ Feedback ($k_{ap}$)\end{tabular} & 
        $k_{ap}$/$\beta_{HK}$/$\beta_{RR}$/$\beta_{SR}$ &
        HK/RR/SR & 
        all  &
        - &
        all &
        -  \\ \hline
        \multirow{5}{*}{\begin{tabular}{c}sRNA-tuned\\ autorepressor\end{tabular}} &$u_1$  $u_2$  & \multirow{2}{*}{all} & \multirow{2}{*}{all} & - & \multirow{4}{*}{all} &  \multirow{4}{*}{-} \\ \cline{2-2}
        & \multirow{7}{*}{$\gamma_R$} & &   &   &   &    \\  \cline{3-5}
        &  & $s,c,T/c,T$ &  \begin{tabular}{c} $\alpha_t$ $\alpha_L$ $n_T$ $K_1$ \\ $\beta_T$  $\delta_t$ $\delta_s$ $\delta_c$ $\delta_T$ \end{tabular} & $K_T$ $K_{u_1}$ $n_{u_1}$ $u_1$ &    &    \\  \cline{3-7}
        &  & $t,T$ &  \begin{tabular}{c} $\alpha_t$ $\alpha_L$ $n_T$ $K_1$ \\ $\beta_T$  $\delta_t$ $\delta_s$ $\delta_c$ $\delta_T$ \end{tabular} & $K_T$ $K_{u_1}$ $n_{u_1}$ $u_1$ & \multirow{3}{*}{t,s,T}    & \multirow{3}{*}{c}   \\ \cline{3-5}
        &  & $s,T$ &  \begin{tabular}{c} $\alpha_t$ $\alpha_L$ $n_T$ $K_1$ \\ $\beta_T$  $\delta_t$ $\delta_s$ $\delta_T$ \end{tabular} & $K_T$ $K_{u_1}$ $n_{u_1}$ $u_1$ $\delta_c$ &    &     \\   \hline
        
        \multirow{13}{*}{Closed-loop sRNA} &
        \multirow{9}{*}{$u_2$ $u_3$} & $R$ & \begin{tabular}{c} $K_{u_2}$ $K_{u_3}$ $K_{R}$ \\ $\delta_R$ $\delta_r$ $\delta_s$\end{tabular} & $\alpha^{*}_r$ $\alpha^{*}_X$ $K_2$ $\delta_c$ $\beta_R$ & R     & r s c \\\cline{3-7}
        & & \multirow{2}{*}{$r$/$s$/$r$  $s$}   & \multirow{2}{*}{\begin{tabular}{c}$\alpha^{*}_r$ $\alpha^{*}_X$ $K_{u_2}$ $K_{u_3}$\\ $K_2$ $\delta_R$ $\delta_r$ $\delta_s$  \end{tabular}}   & \multirow{2}{*}{$K_{R}$ $\delta_c$ $\beta_R$}       & \multirow{2}{*}{r s}   & \multirow{2}{*}{c R}   \\
        & &         &   &     &   &   \\\cline{3-7}
        & & \multirow{2}{*}{\begin{tabular}{c}$c$/$r$ $c$/$s$ $c$/  \\ $r$ $s$ $c$ \end{tabular}}  & \multirow{2}{*}{\begin{tabular}{c}$\alpha^{*}_r$ $\alpha^{*}_X$ $K_{u_2}$ $K_{u_3}$\\ $K_2$  $\delta_R$ $\delta_r$ $\delta_s$ $\delta_c$ \end{tabular}} & \multirow{2}{*}{$K_{R}$ $\beta_R$}          & \multirow{2}{*}{r s c} & \multirow{2}{*}{R} \\
        & &         &  &     &  &      \\\cline{3-7}
        
        & & \multirow{2}{*}{$R$ $r$/$R$ $s$/$R$ $r$ $s$} & \multirow{2}{*}{\begin{tabular}{c}$\alpha^{*}_r$ $\alpha^{*}_X$ $K_{u_2}$ $K_{u_3}$\\ $K_2$ $K_{R}$ $\delta_R$ $\delta_r$ $\delta_s$ $\beta_R$ \end{tabular}} & \multirow{2}{*}{$\delta_c$}          & \multirow{2}{*}{R r s} & \multirow{2}{*}{c}  \\ 
        & &       &  &     &  &    \\\cline{3-7}
        & & $R$ $c$        & all                             & -           & all   & -     \\\cline{2-7}
        & \multirow{4}{*}{$\gamma_X^*$ $\gamma_R^*$} & $R$/$R$ $r$/$R$ $s$/$R$ $r$ $s$ &  $K_2$ $\delta_R$ $\delta_r$ $\delta_s$ $\beta_R$ &  $\delta_c$ & R r s &  c \\ \cline{3-7}
        & & $r$/$s$/$r$ $s$          & $K_2$ $\delta_r$ $\delta_s$  & $\delta_R$ $\delta_c$ $\beta_R$ & r s & c R \\ \cline{3-7}
        & & $c$/$r$ $c$/$s$ $c$/$r$ $s$ $c$          & $K_2$ $\delta_r$ $\delta_s$  $\delta_c$ & $\delta_R$ $\beta_R$ & r s c & R \\ \cline{3-7}
        & & $R$ $c$        & all                             & -           & all   & -     \\\hline
    \end{tabular}
\end{center}
\end{table*}

%%%%%%%%%%%%%%%%%%%%%%%%%%%%%%%%%%%%%%%%%%%%%%%%%%%%%%%%%%%%%%%%%%
\section{Discussion}\label{sec:discussion}

In this paper we have demonstrated the use of symbolic computation to analyse structural properties of synthetic biological circuits. 
Our analyses have shed light on how the availability of output measurements affects parameter identifiability. 
While certain outputs yield identifiable and observable models, others achieve only partial identifiability.  
In contrast, all systems were found to be accessible, as one could expect; our analyses did not find any unforeseen deficiencies. 

While the structural identifiability of a number of synthetic biology models had already been analysed by \cite{haus2023structural}, here we have considered a different set of models. In regard to accessibility and controllability, to the best of our knowledge the results reported in this paper represent the first systematic study of these properties in synthetic biology circuits.

An additional contribution of this work is the implementation of the methods in open source software toolboxes. The main novelty is the development of a Python version of NLcontrollability, which we had previously made available in MATLAB. Together with STRIKE-GOLDD and STRIKEpy, these tools provide implementations of the tests reported here in MATLAB and Python. % environments.

It should be noted that some models could only be analysed under assumptions that render them affine in the inputs, which is a requirement for the application of these accessibility and controllability tests.
Furthermore, the results should be taken as an initial approximation to the properties being studied: our analyses adopt a \textit{structural} viewpoint; they do not consider \textit{practical} limitations that can affect their numerical versions. In particular, the analyses assume that the inputs are continuous, time-varying, and sufficiently exciting; in contrast, in real applications inputs may be restricted to e.g. constant or piecewise constant functions. Taking those limitations into account requires a different set of methods.

%%%%%%%%%%%%%%%%%%%%%%%%%%%%%%%%%%%%%%%%%%%%%%%%%%%%%%%%%%%%%%%%%%
\section*{Data and code availability statement}

All the code developed and used for the analyses reported in this paper is available at:
\url{https://github.com/afvillaverde/NLcontrollability} (for accessibility and controllability) and \url{https://github.com/afvillaverde/strike-goldd} (for structural identifiability and observability). The models analysed in this paper %have been implemented in the above toolboxes, and their files 
can be found in the corresponding \texttt{models} folders.

%%%%%%%%%%%%%%%%%%%%%%%%%%%%%%%%%%%%%

\end{document}